\documentstyle[epsfig]{aipproc}

\begin{document}
\begin{flushright}
  \begin{tabular}[t]{l} 
  KEK-TH-745\\
  February, 2001
 \end{tabular}
 \end{flushright}
\vspace*{0.5cm}
\title{Higgs and SUSY Higgs bosons at future linear collider
\footnote{
Talk given at Linear Collider Workshop 2000 - LCWS 2000,
October 24 -28, 2000, Fermilab.
}}

\author{Yasuhiro Okada
\footnote{Email: yasuhiro.okada@kek.jp}
}
\address{Institute of Particle and Nuclear Studies, KEK\\
Oho 1-1, Tsukuba, Ibaraki 305-0801, Japan}

\maketitle

\begin{abstract}
Theoretical overview on phenomenology of Higgs boson and SUSY Higgs boson  
at a future linear collider experiment is given as a 
Higgs and SUSY Higgs working group summary report for LCWS 2000.
\end{abstract}

\section*{Introduction}
%
%
%
%
Exploring the Higgs sector is one of main motivations for constructing 
an $e^+e^-$ linear collider (LC). At the future LC with the center of
mass energy of 300 - 500 GeV, we can expect to produce $10^4$ - $10^5$
Higgs bosons with an integrated luminosity of 100 -1000 fb$^{-1}$ if the 
Higgs boson mass is 100 -200 GeV. For such a relatively light Higgs boson, 
LC can play a role of a Higgs factory and provide qualitatively different 
information from hadron machines, even when the Higgs particle is
discovered at Tevatron or LHC experiment before the start of the
LC experiment.

The purpose of the Higgs factory is to study various properties of the 
Higgs particle. From the production cross section in the 
$e^+e^- \rightarrow Z^* \rightarrow Z H$ process the Higgs coupling to 
two gauge bosons can be deduced. This coupling constant is closely related to 
the mass-generation mechanism of the gauge boson, because both the 
Higgs-two-gauge-boson coupling and the gauge-boson mass term originate
from the same term in the standard model (SM) Lagrangian. By measuring this
coupling constant, we can test whether or
not the discovered scalar particle really plays a role of the Higgs boson
associated with the electroweak symmetry breaking. Measurement of the 
Higgs boson coupling to fermions is important to understand the 
fermion mass generation mechanism. Eventually, we would like to
reconstruct the Higgs potential itself by measuring three-point
as well as four-point Higgs boson self-coupling constants. These are
fundamental questions of particle physics, and the $e^+e^-$ LC
is an ideal place to answer these questions.

In the Higgs and SUSY Higgs session of this workshop, most of contributions
are related to the above questions. Firstly, how well can various coupling 
constants be determined in the future  $e^+e^-$ and $\gamma \gamma$ 
colliders ? With expected accuracy on those coupling measurements, 
is it possible to distinguish various models such as
the minimal SM, the two Higgs doublet model (2HDM) and the minimal
supersymmetric standard model (MSSM)?  Within a specific model,
for example in the MSSM, how well would various supersymmetric
(SUSY) parameters be determined?
Since the first question was mainly addressed by the
experimental summary talk of this session \cite{batt01}, more theoretical
aspects are discussed here. In the following, a short introduction
to theoretical considerations on the Higgs boson mass is given first,
and then the MSSM Higgs sector and the measurement of anomalous
$ZZH$  and $Z \gamma H$ coupling constants are discussed. Finally,
the Higgs physics at a $\gamma \gamma$ collider is briefly mentioned.

\section*{Higgs Boson Mass}
The most fundamental parameter of the Higgs sector is the mass 
of the Higgs boson. Because the role of the Higgs field is to give masses
to elementary particles through electroweak symmetry breaking, the mass
of the Higgs boson itself reflects the self-interaction of the Higgs 
field. In the minimal SM with one Higgs doublet field, the Higgs boson mass
$(m_H)$ is given by $m_H = \sqrt{2\lambda} v$, where $\lambda$ is 
the self-coupling constant of the Higgs potential, $V=-\mu^2|H|^2 
+ \lambda|H|^4$, and $v\simeq 246$ GeV. This formula implies that
the mass of the Higgs boson is closely related to the strength of
interaction responsible for the electroweak symmetry breaking
and its determination provides us an important clue on the possible 
mechanism of the electroweak symmetry breaking. 

\begin{figure}[b!] 
\centerline{\epsfig{file=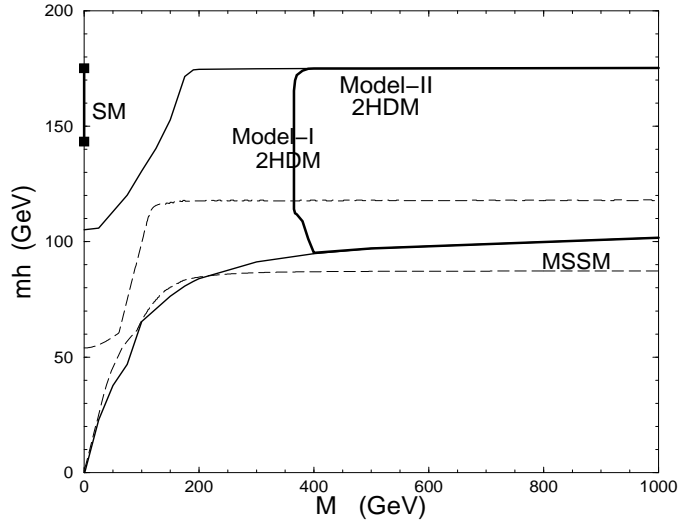,height=3.5in,width=2.8in,angle=-90}}
\vspace{10pt}
\caption{The upper and lower bounds of the lightest CP-even Higgs boson mass
in the type I and II 2HDM for the cutoff scale $\Lambda = 10^{19}$ GeV. 
$M$ is the soft-breaking mass of the discrete symmetry and the heavy Higgs 
boson masses become approximately $M$ for the large $M$ limit.(In the MSSM,
$M = m_A$.) For comparison, the SM Higgs mass bound obtained from one-loop 
renormalization group equations and the lightest 
CP-even Higgs boson mass in the MSSM for $m_{stop}=1$ TeV are shown.}
\label{fig1}
\end{figure}
More precisely, we can derive theoretical upper and lower bounds of the 
Higgs boson mass, if we take a particular model. In the minimal SM,  
the allowed range is 135 GeV $\lesssim m_h \lesssim$ 180 GeV, 
if we require that 
the theory is valid up to the Planck scale, $M_{pl} \sim 10^{19}$ GeV. The 
self coupling constant $\lambda$ blows up below the Planck scale for a 
larger Higgs boson mass than the upper-bound, whereas $\lambda$ turns to 
a negative value and the vacuum stability is not guaranteed if the Higgs 
boson mass is less than
the lower-bound. Similar upper and lower bounds can be obtained for the
2HDM $\cite{kane99}$ and the Zee model of neutrino mass generation
$\cite{zee00}$ as a function of the cutoff scale. 
In these cases the allowed range of the mass for the lightest CP-even Higgs 
boson is 100 GeV $\lesssim m_h \lesssim$ 180 GeV for  
$\Lambda = 10^{19}$ GeV in the
decoupling case where only one SM-like Higgs boson is light compared to
other physical state of the Higgs particles. The mass bounds of the 
lightest CP-even Higgs boson in the 2HDM are shown in Figure \ref{fig1}.

In the MSSM, the upper-bound on the the lightest CP-even Higgs boson mass
$(m_h)$ is determined without reference to the cut-off scale of the theory.
This is because the self-coupling of the Higgs field is completely 
determined by the gauge coupling constants at the tree level. 
Taking into account the top and stop one-loop corrections,  
$m_h$ is given by 
\begin{equation}
m_h^2 \leq m_Z^2 \cos^2{2\beta}+\frac{3}{2\pi^2}\frac{m_t^4}{v^2}
ln{\frac{m_{stop}^2}{{m_t}^2}},
\end{equation}
where $\tan{\beta}\equiv <H_2^0>/<H_1^0>$ is the ratio of the vacuum
expectation values of two Higgs fields\cite{okad91}. 
In the above expression, we have 
assumed that the left- and right-handed stop squarks have the same mass
and there is no mixing among two states. In the literature more precise
calculation is available, and it is concluded that $m_h$ is bounded
by about 130 GeV, even if we take the stop mass to be a few TeV. Because 
the expected mass range is somewhat lower than that derived under assumption 
that the minimal SM is valid up to the Planck scale, the discovery of
the Higgs boson around 120 GeV would be a strong indication of the MSSM.

In extended versions of SUSY model, the upper-bound of the lightest 
CP-even Higgs boson can be determined only if we require that any of
dimensionless coupling constants of the model does not blow up below some 
cut-off scale. For the SUSY model with an extra gauge singlet Higgs field,
the bound is about 150 GeV, which is a slightly larger than the upper-bound 
for the MSSM case. Because there is a new tree level contribution
to the Higgs mass formula, the maximum value corresponds to a lower 
value of $\tan{\beta}$, which is quite different from the MSSM case where
the Higgs mass becomes larger for large $\tan{\beta}$. In Ref.\cite{espi98}
the upper-bound of the lightest CP-even Higgs boson was calculated for SUSY 
models with gauge-singlet or gauge-triplet Higgs field and the maximal
possible value was studied in those extensions of the MSSM. It was 
concluded that the mass bound can be as large as 210 GeV for a specific 
type model with a triplet-Higgs field for a stop mass of 1 TeV. The mass 
bound was also studied for the SUSY model with extra matter fields. In this 
model the upper-bound becomes larger due to loop corrections of extra matter
multiplets. If the extra fields have $\bar{5}+10+5+\bar{10}$ representations
in SU(5) GUT symmetry, the maximum value of the lightest CP-even Higgs 
boson mass becomes 180 GeV for the case that the squark mass is 1 TeV 
\cite{moro92}.   

\subsubsection*{Detectability of at least one Higgs boson at LC} 
Because the above upper-bound of the lightest CP-even Higgs boson is 
at most about 200 GeV, at least one of the Higgs boson is kinematically 
accessible in any type of SUSY models for the $e^+e^-$ LC with $\sqrt{s}=$ 500 
GeV. It is, however, important to know whether such a Higgs boson can be
discovered for reasonable integrated luminosity, because the production 
cross section and decay modes depends on details of models. In the MSSM,
it was shown that at least one of two CP-even Higgs bosons can be 
detectable through $e^+e^- \rightarrow Zh_i$ process with relatively low
integrated luminosity $(\sim 10 $fb$^{-1})$. In the model with a singlet Higgs 
field there are three CP-even Higgs bosons, and if the lightest one becomes
singlet-dominated, its mass bound alone does not guarantee the discovery of
the Higgs boson because its coupling to gauge bosons is very suppressed. 
Even in such a case we can show that at least one Higgs boson with
a sizable SM Higgs component exists below or around the upper-bound
of the lightest CP-even Higgs boson. In fact, we can calculate
a minimal production cross section which means that at least one 
of three CP-even Higgs bosons has a larger production cross section than 
that value in the $e^+e^- \rightarrow Zh_i$ process. The minimal cross
section turns out to be one third of the SM Higgs production cross
section evaluated at the upper-bound value. This production cross section 
is typically 20 - 50 fb for $\sqrt{s}=$ 300 - 500 GeV and large enough 
to guarantee the discovery of at least one Higgs boson independently of its
decay modes\cite{kamo94}. Furthermore, in more general SUSY models with 
arbitrary number of the Higgs fields, it was shown that the Higgs signal 
is observable at $e^+e^-$ LC with $\sqrt{s}=$ 500 GeV and integrated 
luminosity of 500 fb$^{-1}$ \cite{espi99}. This is true even if the Higgs 
bosons decay to invisible modes because we can measure a recoil mass 
distribution at the LC experiment.
 
\subsubsection*{Precision electroweak measurements v.s. the Higgs boson mass} 
Since present electroweak measurements are precise enough to be sensitive to 
virtual loop effects, a useful constraint on the Higgs boson mass is obtained
in the minimal SM. Taking account of the direct measurement of 
the $W$ boson and top quark mass in  addition to various electroweak data
at LEP and SLD experiments, the 95 \% upper bound on the SM Higgs boson mass
is about 210 GeV\cite{lepwg}. This value means that the Higgs boson is 
detectable in the future LC with $\sqrt{s}=$ 500 GeV. Because this bound is 
obtained within the minimal SM, it is very interesting whether or not this
kind of bound can be derived in other models.

In general 2HDM, unlike the MSSM case, we cannot derive useful mass bound
if we do not consider the condition on the validity of the theory up to 
some cut-off scale. The question whether we can conclude that at least one of 
Higgs boson will be discovered at LC based on present knowledge of the 
electroweak measurements was considered in Ref. \cite{guni00} 
and it was shown that 
the following four conditions could be satisfied simultaneously: (1)
The light Higgs boson (CP-even $(h)$ or CP-odd $(A)$) has no $ZZ/WW$ couplings.
(2) The heavy CP-even Higgs boson $(H)$ is too heavy to be produced
at LC with $\sqrt{s}=$ 500 - 800 GeV. (3) The production cross section for
$e^+e^- \rightarrow t\bar{t}h$ or $b\bar{b}h$ ($t\bar{t}A$ or $b\bar{b}A$) 
are not large enough for discovery in these modes even with extremely
large integrated luminosity (2500 fb$^{-1}$). (4) These parameters
are still reasonably consistent with the present electroweak precision
test. Although no direct signal on the Higgs sector is expected in this
particular parameter space, we can distinguish this model if we improve
the electroweak measurement including the $W$ boson mass determination at 
the Giga-Z option of the LC experiment\cite{guni00,pesk01}.     

\section*{MSSM Higgs sector}
\begin{figure}[b!] 
\centerline{\epsfig{file=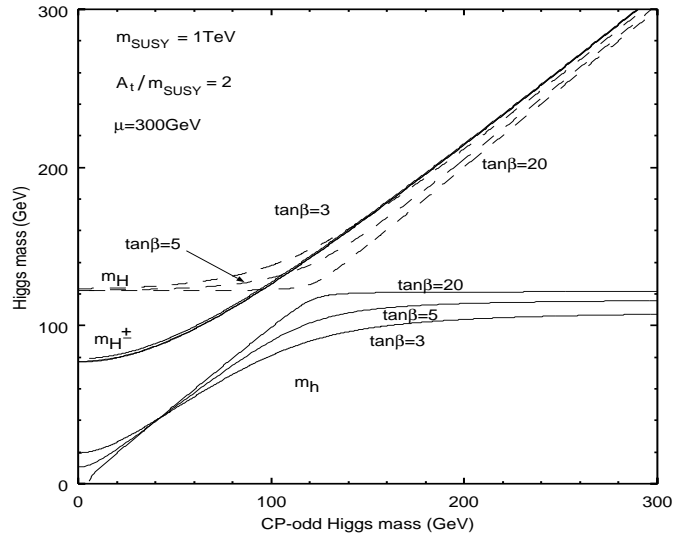,height=2.8in,width=3.5in}}
\vspace{10pt}
\caption{Higgs boson masses as a function of the CP-odd Higgs boson mass 
in MSSM.}
\label{fig2}
\end{figure}
In the MSSM, physical Higgs states are CP-even Higgs bosons $(h, H)$,
a CP-odd Higgs boson $(A)$ and a pair of charged Higgs bosons $(H^\pm$). 
At the tree level, the Higgs potential is parametrized by three mass 
parameters in addition to gauge coupling constants. It is known that the 
radiative corrections, mainly due to top and stop loop diagrams, can give 
important modification to the Higgs potential, so that the lightest CP-even 
boson mass receives a large radiative correction. 
Three dimensionful parameters can be
taken as two vacuum expectation values, in other words $v$ and 
$\tan{\beta}$, and one of Higgs boson masses, which is usually taken as the 
CP-odd Higgs boson mass $(m_A)$. In addition, the Higgs sector depends on
the top and the stop masses through the radiative correction. More
precisely, the parameters appearing in the formula of the radiative correction 
are not just one stop mass, but two stop masses, 
trilinear coupling constant for stop sector $(A_t)$, the higgsino mass 
parameter $\mu$, and sbottom masses, 
etc. (If we use more precise formula, we need to specify more input
parameters.) Once these parameters are specified, we can calculate the mass
and the mixing of the Higgs sector. The Higgs mixing parameter $\alpha$
is defined by
$Re H_1^0=(v \cos{\beta} - h \sin{\alpha} + H \cos{\alpha})/\sqrt{2}$, 
$Re H_2^0=(v \sin{\beta} + h \cos{\alpha} + H \sin{\alpha})/\sqrt{2}$,
where $H_1^0$ and $H_2^0$ are neutral component of two Higgs 
doublet fields. $\alpha$ is a function of independent parameters 
($m_A,\tan{\beta},m_t,m_{stop}$, etc).

The masses of the CP-even and the CP-odd Higgs bosons as well as 
the charged Higgs bosons are shown as a function of $m_A$ in
Fig. \ref{fig2}.
It is important to distinguish two regions in this figure.
Namely, when $m_A$ is much larger than 150 GeV, $H, A$ and $H^{\pm}$
states become approximately degenerate and the mass of $h$ approaches
to its upper-bound value for each $\tan{\beta}$. This limit is
called the decoupling limit. In this limit, $h$ has properties similar 
to the SM Higgs boson, and the coupling of the heavy Higgs bosons to
two gauge-boson states is suppressed. On the other hand, if $m_A$
is less than 150 GeV, the lightest CP-even Higgs boson has a sizable
components of the SM Higgs field and the other doublet field.
Since LEP experiments have already excluded a significant portion
of the parameter space, it is quite possible that the Higgs boson
will be found in the decoupling region. If we discover only one
Higgs boson at either Tevatron or LHC experiments in the mass region 
consistent with the MSSM, precise
measurements of the Higgs particle's properties are important
to distinguish this from the SM Higgs boson, especially in the decoupling
and near-decoupling regions. The future LC is the most suitable place to
perform such precise measurements. In the following subsections, we 
discuss various ways to distinguish the MSSM from other models 
through measurements with respect to the Higgs sector.

\begin{figure}[b!] 
\centerline{\epsfig{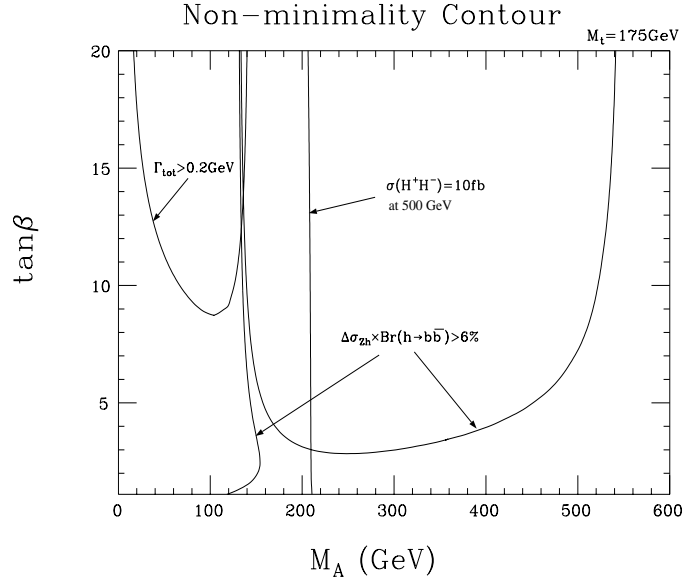}}
\vspace{10pt}
\caption{Parameter space in the MSSM where deviation
from the SM is expected to be observable at $e^+e^-$ LC. 
Three lines corresponds to: 
(1) The total width of the lightest CP even Higgs boson mass 
is larger than 200 MeV, (2) $\sigma (e^+e^- \rightarrow Z h) 
Br (h\rightarrow b \bar{b})$
deviates from the SM value more than  6\%
(3) $\sigma (e^+e^- \rightarrow H^+H^-) > 10$ fb for 
$\sqrt{s}=500$ GeV. This figure is provided by A. Miyamoto.}
\label{fig3}
\end{figure}     
\subsubsection*{Non-minimality of the Higgs sector}
In the production of the light Higgs boson, the number of the 
Higgs signals provide important information on the model behind
the Higgs sector. Because the main decay mode of the light Higgs 
boson is $h \rightarrow b \bar{b}$, $\sigma (e^+e^- \rightarrow
Z h) B(h \rightarrow b \bar{b})$ is a quantity expected to be 
measured precisely.
In Fig. \ref{fig3}, the MSSM parameter region
where $\sigma (e^+e^- \rightarrow Z h) B(h \rightarrow b \bar{b})$ 
differs from the SM prediction by 6\% is shown for $m_t = $175 GeV
and $m_{stop} =$ 1 TeV. Because the statistical accuracy
of this quantity at the first stage of the future LC experiment 
is expected to reach a few \% level, the deviation will be clear
within this parameter region. The contour extends to the line
corresponding to $m_A= 500$ GeV. In this figure
two other lines are drawn . One corresponds to the case in which 
the total width of the Higgs boson is larger than 200 MeV, so that the 
recoil mass distribution in the $ e^+e^- \rightarrow Z h \rightarrow
l^+l^-h$ provides the direct measurement of this quantity. 
The other line is the limit where the charged Higgs boson pair production 
is observable at the LC with $\sqrt{s} =$ 500 GeV.

\subsubsection*{Higgs boson branching ratios v.s. the heavy Higgs boson 
mass } 
\begin{figure}[b!] 
\centerline{\epsfig{file=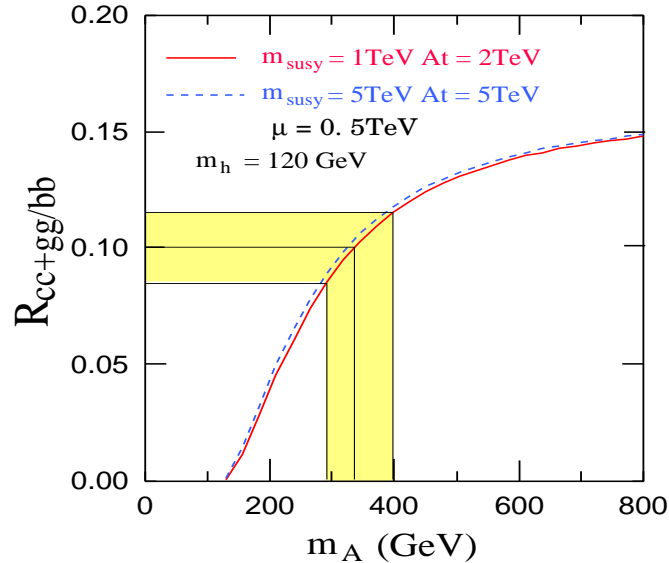,height=3in,width=3.5in}}
\vspace{10pt}
\caption{$R_{cc+gg/bb}\equiv
\protect\frac{(Br(h\to c\bar{c})+Br(h\to gg))}{Br(h\to b\bar{b})}$ is shown
as a function of $m_A$ for different choices of SUSY parameters.
Expected statistical error for 100 fb$^{-1}$ is also shown.}
\label{fig4}
\end{figure}

It was pointed out that the following ratio of the branching ratios
\begin{equation}
R_{cc+gg/bb}=\frac{B(h \rightarrow c \bar{c})+B(h \rightarrow
gg)}{B(h \rightarrow b\bar{b})}
\end{equation}
is a sensitive probe to $m_A$\cite{kamo96}.
In the MSSM, the Higgs sector is the same as
type II 2HDM at the tree level, so that the up(down)-type Yukawa coupling 
constant is associated with $H_2$ $(H_1)$, and $B(h \rightarrow c \bar{c})$
($B(h \rightarrow b \bar{b})$) is proportional to
$\cos^2{\alpha}/\sin^2{\beta}$ $(\sin^2{\alpha}/\cos^2{\beta})$. Because 
$h \rightarrow gg$ is induced predominantly by the internal top quark loop,
the dependence on the angles is the same as the $h \rightarrow c \bar{c}$
mode. As a result, $R_{cc+gg/bb}$ is proportional to $1/(\tan{\beta}
\tan{\alpha})^2$. Taking into account the top and the stop loop 
corrections to the Higgs potential, we can relate these angles to other
input parameters and derive the following approximate formula.
\begin{equation}
R_{cc+gg/bb}\simeq \left( \frac{m_A^2 -m_h^2}{m_A^2+m_Z^2} \right) 
^2 R_{cc+gg/bb}(SM),
\end{equation}
where $R_{cc+gg/bb}(SM)$ is the same ratio evaluated in the SM.
In Fig.\ref{fig4}, $R_{cc+gg/bb}(SM)$ is shown for $m_h=$ 120 GeV. 
($\tan{\beta}$ is solved for each $m_A$ to give this Higgs boson mass.)
This figure shows that the above formula is actually a very good 
approximation and the ratio is almost independent of other parameters like
$m_{stop}$ and $A_t$. The expected statistical error for 100 fb$^{-1}$ 
is also shown\cite{naka96}. We can see that this quantity is useful 
to constrain the mass scale of the heavy Higgs bosons. 

Although the lightest CP-even
Higgs boson is likely to be discovered in the LHC experiment,
whether or not other Higgs states $(H, A, H^\pm)$ can be found depends 
on  the SUSY parameters.
There is a large portion of the parameter space with $m_A \gtrsim
200$ GeV and $2 \lesssim \tan{\beta} \lesssim 10$ where only one
lightest CP-even Higgs boson will be discovered. If this is the case, 
it would be very important that possible range of the heavy Higgs boson
mass could be inferred from branching ratio measurements at the first
stage LC experiment with $\sqrt{s}=$ 300 - 500 GeV, so that we can set
the target energy of the second stage LC experiment.       

\subsubsection*{SUSY loop correction to bottom Yukawa coupling
at large $\tan{\beta}$}
When $\tan{\beta}$ is as large as 50, SUSY loop corrections generate
important contribution to the bottom-Higgs Yukawa coupling\cite{babu99}.
Including one-loop induced coupling the top and bottom Yukawa couplings
with the neutral Higgs fields are given by
\begin{equation}
L_{Yukawa}=y_t\bar{t_L}t_R H_2^0 + y_b\bar{b_L}b_R H_1^0 +
\epsilon_b y_b\bar{b_L}b_R H_2^{0*} + h.c.
\nonumber
\end{equation} 
The $\epsilon_b$ term is induced by loop diagrams with internal 
sbottom-gluino and stop-chargino. Thus the Higgs 
sector becomes effectively general-type 2HDM, not restricted to 
the type-II model. The bottom mass is expressed by  
\begin{equation}
m_b=y_b(1+\epsilon_b\tan{\beta})\frac{v}{\sqrt{2}}\cos{\beta}.
\label{mb}
\end{equation}
Although $\epsilon_b$ is typically $O(10^{-2})$,
the correction to $m_b$ enters with a combination of 
$\epsilon_b\tan{\beta}$ which can be close to $O(1)$ for a large
value of $\tan{\beta}$. 
Precise calculation of various branching ratios was presented
taking account of these corrections\cite{weig00}. 
For example, the ratio of $B(h \rightarrow \tau^+\tau^-)$ and  
$B(h \rightarrow b\bar{b})$ is modified to
\begin{equation}
R_{\tau \tau/bb}\equiv  \frac{B(h \rightarrow \tau^+\tau^-)}
{B(h \rightarrow b\bar{b})}=\left( \frac{1+\epsilon_b \tan{\beta}}
{1-\epsilon_b/\tan{\alpha}} \right)^2 R_{\tau \tau/bb}(SM),
\nonumber
\end{equation}
where $R_{\tau \tau/bb}(SM)$ is the same ratio evaluated in the SM.
$R_{\tau \tau/bb}$ is the same as $R_{\tau \tau/bb}(SM)$ if
the SUSY loop effect to the $b\bar{b}H_2^{0*}$ vertex is negligible. 
We also notice that in the decoupling limit with $m_A \rightarrow \infty$,
$\tan{\alpha}$ is approaching to $-1/\tan{\beta}$, so that the ratio reduces 
to the SM prediction. This is because the properties of the lightest
CP-even Higgs boson become similar to those of the SM Higgs boson,
independent of how the two Higgs doublet fields couple to the fermions. 
In actual evaluation, however, the approach to the asymptotic 
form is slow for large $\tan{\beta}$, so that sizable deviation
from the SM prediction may be observed.

The SUSY vertex correction also modifies the previous discussion on 
constraining $m_A$ from $R_{cc+gg/bb}$. This problem was investigated and 
it was shown that the effect on   $R_{cc+gg/bb}$  is within a few \%
for $\tan{\beta}=$ 10, although the uncertainty is too large for    
$\tan{\beta}=$ 50 \cite{kiyo01}. This means that for the parameter space 
where the LHC experiment would not find the heavy Higgs boson, 
$R_{cc+gg/bb}$ can still give important clue on the heavy Higgs mass scale. 
The theoretical uncertainty to
the branching ratio from input parameters such as $\alpha_s, m_b$ and $m_c$
was also studied, and the error is about 8\% for
$R_{cc+gg/bb}(SM)$ and $R_{\tau \tau/bb}(SM)$ if we know
$\alpha_s$ within 2\%, the $\overline{MS}$ running quark masses, 
$m_b(m_b)$ within 3\% and $m_c(m_c)$ within 5\%. 
The mass determination at this level from lattice gauge theory is
very important for Higgs physics. 

Once we measure $B(h \rightarrow b\bar{b})$, $B(h \rightarrow c\bar{c})$
and $B(h \rightarrow \tau^+ \tau^-)$ with enough precision, we can determine
the correction factor $\Delta_b\equiv \epsilon_b  \tan{\beta}$
\cite{loga00}.
If x(y) is defined as 
$B(h \rightarrow b\bar{b})/B(h \rightarrow \tau^+ \tau^-)$ 
($B(h \rightarrow c\bar{c})/B(h \rightarrow \tau^+ \tau^-)$)
normalized by the corresponding SM values, we get
$\Delta_b=(1-\sqrt{x})/(\sqrt{x}-\sqrt{y}).$
Determination of $\Delta_b$ is very interesting
because non-zero value provides a hint for the SUSY loop effect on
the vertex correction.

The vertex correction also modifies the production and the decay 
of the heavy Higgs bosons\cite{kold00}. 
Main effects on the coupling of heavy Higgs 
boson to the fermions is due to the correction to the relation between
the bottom mass and the original bottom Yukawa coupling constant shown 
in Eq.(\ref{mb}).
In SO(10)-like GUT model with Yukawa unification, for example,
$\Delta_b$ should be -25\% to -18\% and the bottom coupling constants to 
$A$ and $H$ become effectively larger by the same amount. As a result, the
production cross sections of $e^+e^- \rightarrow A b \bar{b}$, $H b \bar{b}$
below the threshold of the $A$ $H$ associated production is increased,
and the $\tau^+ \tau^-$  and $c \bar{c}$ branching ratios of the heavy Higgs
boson are suppressed. 

\subsubsection*{Determination of $\tan{\beta}$ from heavy Higgs boson process}
In SUSY models $\tan{\beta}$ is a very important parameter.
Although it is defined as a ratio of two vacuum expectation values, 
this parameter appears in various couplings of the MSSM Lagrangian.
The consistency check of $\tan{\beta}$ determined in the Higgs sector,
the chargino-neutralino sector as well as the squark-slepton sector
will be interesting because these relationships result from SUSY
invariance. 

In the MSSM, the possible range of $\tan{\beta}$ is 1 - 70 from the 
requirement that the Yukawa coupling constants do not blow up below the
Planck scale. The range of $\tan{\beta} \lesssim 2$ is excluded from the
LEP II Higgs boson search for reasonable values of stop masses and mixing
parameters. Direct determination of $\tan{\beta}$
is a challenging task for future LC experiment.

One of best ways to determine $\tan{\beta}$ is to study heavy Higgs boson 
production and decays, because the branching ratios and production cross
sections depend on $\tan{\beta}$ significantly. For the branching ratios,
$A, H \rightarrow t \bar{t}$ and $H^- \rightarrow b \bar{t}$ are dominant
for small  $\tan{\beta}$ as long as these modes are kinematically allowed,
whereas $A, H \rightarrow b \bar{b}, \tau^+ \tau^- $ and  
$H^- \rightarrow \tau \bar{\nu}$ becomes important for larger values of
$\tan{\beta}$. For the production processes, $e^+e^- \rightarrow b \bar{b} 
A$, $b \bar{b} H$ and $e^+e^- \rightarrow t \bar{b} H^-$ are enhanced for large
$\tan{\beta}$. The question to what extent $\tan{\beta}$ can be determined
at future LC was addressed in Ref. \cite{feng97}. It was concluded that 
$\tan{\beta}$ would be well constrained in the intermediate values 
of $\tan{\beta}$ $( 3 \lesssim \tan{\beta} \lesssim 10)$. In addition, 
$e^+e^- \rightarrow t \bar{b} H^-$ production cross section is enhanced 
for a large value of $\tan{\beta} ( \gtrsim 50)$. 

In this workshop, a new analysis was reported where systematic study
was performed using $e^+e^- \rightarrow H b \bar{b}, H t \bar{t}$,
$A b \bar{b}$, $A t \bar{t}$ process at LC with $\sqrt{s} = 500 - 1000$
GeV\cite{jian00}. Within the MSSM scenario, $\tan{\beta}$ will be well 
constrained for $\tan{\beta} \lesssim 10$. Typical error is $\pm 2$ for
$\tan{\beta} = 3$ and  $\pm 4$ for $\tan{\beta} = 10$. If $\tan{\beta}$ 
is determined from the Higgs sector we can use this value as an input
parameter for various other SUSY processes such as chargino production 
and stau decays, so that we can perform quantitative tests of SUSY
Lagrangian.

\subsubsection*{Single charged Higgs boson production} 
Main production process of the charged Higgs boson at LC
is $e^+e^- \rightarrow H^+ H^-$, so that the mass reach
is kinematically restricted to $\sqrt{s}/2$. A single charged Higgs boson 
production is potentially important because the search for a heavier charged 
Higgs boson is possible.

For this purpose, a systematic investigation of single charged Higgs boson
production processes at $e^+e^-$ LC has been done\cite{kane00}.
Production cross sections for 14 processes were calculated in the MSSM.
These are $e^+e^- \rightarrow \tau^+ \nu H^-$, $t\bar{b}H^-$, $W^+H^-$,
$e^+ \nu H^-$, $Z^0 W^+H^-$, $h W^+H^-$, $H W^+H^-$, $A W^+H^-$, 
$e^+e^- W^+H^-$, $\nu \bar{\nu}W^+H^-$, $e^+ \nu Z^0 H^-$, 
$e^+ \nu h H^-$, $e^+ \nu H H^-$, and $e^+ \nu A H^-$. 
Among these processes,  $e^+e^- \rightarrow
W^+H^-$, $e^+ \nu H^-$ are one-loop induced processes. Generally, the 
production cross sections are very suppressed $( < 10^{-2} $fb$^{-1})$
below the $H^+H^-$ pair production threshold, but the two
processes are found to be promising. Namely, $e^+e^- \rightarrow
\tau^+ \nu H^-$ for large $\tan{\beta}$ case and $e^+e^- \rightarrow W^+H^-$ 
for small $\tan{\beta}$ case have production cross sections of $0.01
- 1 $fb$^{-1}$ for $\sqrt{s}= 500$ GeV, and therefore may
be observable with integrated luminosity of 500 fb$^{-1}$.    

\section*{Anomalous coupling measurement}
At $e^+e^-$ LC experiment, a great number of light Higgs bosons are 
expected to be produced through the $e^+e^- \rightarrow ZH \rightarrow 
f\bar{f}H$ process. We can put constraints on anomalous coupling constants 
for the $ZZH$ and $Z \gamma H$ vertexes, because these coupling constants 
generate additional contribution to the $ZH$ production process through 
interference with the SM process \cite{kamo00}. Up to dimension-five
operators these anomalous couplings are given by
\begin{eqnarray}
L_{eff} &=& (1+a_z)\frac{g_Z m_Z}{2} H Z_{\mu} Z^{\mu}\nonumber\\
& & +\frac{g_Z}{m_Z}\sum_{V=\gamma,Z}[ b_V H Z_{\mu \nu} Z^{\mu \nu}
+c_V (\partial_{\mu} Z_{\nu} - \partial_{\nu} Z_{\mu}) V^{\mu \nu}
+\tilde{b}_V H Z_{\mu \nu} \tilde{V}^{\mu \nu}]
\end{eqnarray}    
where $V_{\mu} = A_{\mu}$ or $Z_{\mu}$ and $V_{\mu \nu}=
\partial_{\mu} V_{\nu} - \partial_{\nu} V_{\mu}, \tilde{V}_{\mu \nu}=
\epsilon_{\mu \nu \alpha \beta} V^{\alpha \beta}$. For $\sqrt{s}=500$ GeV 
and integrated luminosity of 300 fb$^{-1}$, expected constraints on these 
coupling constants
were studied for $m_H = $115 GeV. In particular, effects of helicity 
measurements of $\tau$ lepton ($\epsilon_{\tau}= 50$\%), 
charge identification of the bottom quark ($\epsilon_{b}= 60$\%)
as well as beam polarization ($|P_{e^-}| $ = 80 \%, $|P_{e^+}| $ = 45 \%) 
were investigated. These options turns out to be useful to improve the 
measurements of anomalous coupling constants. With these options the 
above parameters can be constrained to the level of $10^{-4}$.
In particular, the improvement is significant for the $Z\gamma H$ coupling
measurement. This is because the interference term between the SM
amplitude and the amplitude with the $Z \gamma H$ couplings is 
suppressed without these options due to the almost axial-vector nature of
the $e^+e^- Z$ coupling in the SM.

\section*{Photon-photon collider}
In the photon-photon option of the LC experiment, we can directly measure 
$\Gamma(H \rightarrow \gamma \gamma) B(H \rightarrow b \bar{b})$ by the
s-channel Higgs boson production process. Assuming that 
$B(H \rightarrow b \bar{b})$
will be already precisely known from the $e^+ e^-$ LC
experiment, we can obtain the two photon partial width of the Higgs boson.
Combined with the information on $B(H \rightarrow \gamma \gamma)$, the
total decay width will be derived. Because  
$\Gamma(H \rightarrow \gamma \gamma)$ is induced by heavy particle loops,
accurate determination of $\Gamma(H \rightarrow \gamma \gamma)$ is very 
interesting as a possible window to new physics.

In order to estimate the accuracy, we have to evaluate rates of the signal 
and background processes in the SM. In this workshop, recent developments
on the radiative corrections for these processes were reported
\cite{yako00,mell00}. 
In the SM background process, $e^+e^- \rightarrow b \bar{b}$, the Born
cross section for $J_z =0$ channel is suppressed by $m_b^2/s$
compared to that for  $J_z =\pm 2$ channel, so that we can use 
polarization of initial photons to suppress the background. On the other hand,
there is a large QCD correction to this process and careful treatment 
of higher order corrections is necessary. The radiative corrections to
the SM signal and background processes are under control and it was reported 
that the measurement of two photon decay width with a statistical accuracy 
of 1.4\% would be possible for a Higgs boson mass of 115 GeV using the TESLA  
parameters\cite{mell00}. 

One example of possible role in identifying a new physics from
the two photon partial width of the Higgs boson was discussed\cite{ginz01}. 
In 2HDM, it may be possible that only a light Higgs boson is found
at LHC and $e^+e^-$ LC experiments and its couplings to gauge boson and 
fermions are found to be consistent with the SM Higgs boson within the
expected accuracy of these experiments. Even in such a situation, the  
deviation of $\Gamma(H \rightarrow \gamma \gamma)$ from the SM value can be 
as large as 10\%, so that the two-photon decay width measurement 
may be able to provide the first hint for non-mimimality of the Higgs sector.

\section*{Summary}
We have discussed various examples in which studies of the Higgs sector
at future LC experiment provide signals of physics beyond the SM. 
If in fact a light Higgs boson is found at Tevatron, LHC or LC 
experiments, the role of LC will be the Higgs factory. By precise
determination of production cross section and branching fractions, 
we can test whether the observed particle corresponds to the SM Higgs 
boson or the MSSM Higgs boson. Within the MSSM we can put a 
constraint on the SUSY parameters. If kinematically allowed, heavy charged 
and neutral Higgs bosons may be produced, and their properties 
are useful to identify models. The photon-photon option is also 
useful to search for new physics through loop effects in the two
photon decay width of the Higgs boson. These measurements are 
important to establish or distinguish various candidates of
models such as the minimal SM, MSSM, 2HDM or some other 
possibilities which may be highlighted in future.\\  

This work was supported by the Grant-in-Aid of the Ministry of Education,
Science, Sports and Culture, Government of Japan (No.\ 09640381),
Priority area ``Supersymmetry and Unified Theory of Elementary Particles''
(No.\ 707), and ``Physics of CP Violation'' (No.\ 09246105).

\end{document}